\documentclass[usenatbib]{mn2e}

\pdfoutput=1

\usepackage[pdftex]{graphicx}
\usepackage{mathptmx}
\usepackage{fixltx2e}
\usepackage{color}

\newcommand{\Mpc}{\rm\thinspace Mpc}
\newcommand{\kpc}{\rm\thinspace kpc}

\newcommand{\km}{\rm\thinspace km}

\newcommand{\cm}{\rm\thinspace cm}

\newcommand{\pcmcu}{\hbox{$\cm^{-3}\,$}}

\newcommand{\yr}{\rm\thinspace yr}

\newcommand{\s}{\rm\thinspace s}

\newcommand{\Msun}{\hbox{$\rm\thinspace M_{\odot}$}}

\newcommand{\Msunpyr}{\hbox{$\Msun\yr^{-1}\,$}}
\newcommand{\keV}{\rm\thinspace keV}

\newcommand{\erg}{\rm\thinspace erg}

\newcommand{\ergpcmsqps}{\hbox{$\erg\cm^{-2}\s^{-1}\,$}}

\newcommand{\ergps}{\hbox{$\erg\s^{-1}\,$}}

\newcommand{\kmps}{\hbox{$\km\s^{-1}\,$}}

\newcommand{\kmpspMpc}{\hbox{$\kmps\Mpc^{-1}$}}
\newcommand{\Zsun}{\hbox{$\thinspace \mathrm{Z}_{\odot}$}}

\newcommand{\psqcm}{\hbox{$\cm^{-2}\,$}}
\newcommand{\pcmsq}{\hbox{$\cm^{-2}\,$}}

\begin{document}

\title[The Coma cluster brightest galaxy X-ray coronae]
{The X-ray coronae of the two brightest galaxies in the Coma
  cluster}

\author
[J.~S. Sanders et al]
{
  \begin{minipage}[b]{\linewidth}
    \begin{flushleft}
      J.~S. Sanders$^1$, A.~C. Fabian$^2$, M. Sun$^3$, 
      E.~Churazov$^{4,5}$,
      A.~Simionescu$^{6,7,8}$,
      S.~A.~Walker$^{2}$ and
      N.~Werner$^{6,7}$
    \end{flushleft}
  \end{minipage}
  \\
  $^1$ Max-Planck-Institut f\"ur extraterrestrische Physik,
  Giessenbachstrasse 1, 85748
  Garching, Germany\\
  $^2$ Institute of Astronomy, Madingley Road, Cambridge. CB3 0HA\\
  $^3$ Department of Physics, University of Alabama in Huntsville,
  Huntsville, AL 35899, USA\\
  $^4$ Max-Planck-Institut f\"ur Astrophysik,
  Karl-Schwarzschild-Strasse 1,
  85748 Garching, Germany\\
  $^5$ Space Research Institute (IKI), Profsoyuznaya 84/32, Moscow 117997, Russia\\
  $^6$ KIPAC, Stanford University, 452 Lomita Mall, Stanford, CA 94305, USA\\
  $^7$ Department of Physics, Stanford University, 382 Via Pueblo
  Mall, Stanford, CA 94305-4060, USA\\
  $^8$ Institute of Space and Astronautical Science (ISAS), JAXA,
  3-1-1 Yoshinodai, Chuo-ku, Sagamihara, Kanagawa 252-5210, Japan }
\maketitle

\begin{abstract}
  We use deep \emph{Chandra X-ray Observatory} observations to examine
  the coronae of the two brightest cluster galaxies in the Coma
  cluster of galaxies, NGC\,4874 and NGC\,4889. We find that NGC\,4889
  hosts a central depression in X-ray surface brightness consistent
  with a cavity or pair of cavities of radius 0.6~kpc. If the central
  cavity is associated with an AGN outburst and contains relativistic
  material, its enthalpy should be around $5 \times 10^{55} \erg$.
  The implied heating power of this cavity would be around an order of
  magnitude larger than the energy lost by X-ray emission.  It would
  be the smallest and youngest known cavity in a brightest cluster
  galaxy and the lack of over pressuring implies heating is still
  gentle.  In contrast, NGC\,4874 does not show any evidence for
  cavities, although it hosts a well-known wide-angle-tail radio
  source which is visible outside the region occupied by the X-ray
  corona. These two galaxies show that AGN feedback can behave in
  varied ways in the same cluster environment.
\end{abstract}


\section{Introduction}
Contrary to theoretical expectations, it was found that the two
dominant central galaxies in the Coma cluster, NGC\,4874 and
NGC\,4889, showed extended X-ray emission from their interstellar
medium \citep{Vikhlinin01}. This material was found to be confined by
the pressure of the surrounding intracluster medium (ICM).
\cite{Sun07} examined a sample of galaxies in clusters, finding these
embedded coronae to be relatively common ($>60$\% in galaxies with
$L_{K_s} > 2L_{*}$), but with decreasing likelihood in hotter
environments. Although these galactic coronae are smaller than those
found outside groups and clusters, their high abundance shows that
they are able to survive for long periods in the harsh intracluster
environment. For example, heat conduction must be highly suppressed
between these galaxies and the surrounding ICM. For these coronae to
be common (i.e. long lasting) they must contain a heat source able to
combat the radiative energy loss due to the emission of the observed
X-rays.

The two dominant galaxies in Coma, NGC\,4889 and NGC\,4874, show a
large line-of-sight velocity difference, $\sim 700 \kmps$, between
them \citep{FitchettWebster87}. The presence of these galaxies is
likely due to a recent cluster merger. The X-ray emission shows that
the cluster is in an unrelaxed state \citep{Briel92}. One or both of
the central galaxies cannot be at rest in the cluster potential or
ICM. Both of the galaxies are associated with galaxy subgroups
\citep{Adami05}.

Here we examine new deep \emph{Chandra} observations of the two
galaxies. These observations represent some of the deepest
observations by \emph{Chandra} of coronae embedded in clusters. We
leave examination of the other coronae in Coma to a later work. The
Coma cluster lies at a redshift of 0.0231 \citep{StrubleRood99},
implying 1 arcsec on the sky corresponds to 0.47 kpc if $H_0 = 70
\kmpspMpc$.

\section{Data analysis}
We processed the \emph{Chandra} Coma datasets as detailed in
\cite{SandersComa13}. The total exposure time of the observations is
546~ks (see Table \ref{tab:chandra}). The event files were reprocessed
to take advantage of VFAINT grading, if appropriate, and reprojected
to the coordinate system of the 13994 observation. No periods from the
observations were removed as no flares were
observed. Blank-sky-background observations were used to create
background event files, adjusting their exposure time to match the 9
to 12 keV count rate in the observed data. Images, background images
and exposure maps were created for each observation and combined.

\begin{table}
  \caption{\emph{Chandra} datasets examined here. Listed are the
    observation identifier, date of observation, ACIS detector mode
    and exposure.}  \centering
  \begin{tabular}{lllr}
    \hline
    Observation & Date & Mode & Exposure (ks) \\
    \hline
    555   & 1999-11-04 & FAINT & 8.7 \\
    1112  & 1999-11-04 & FAINT & 9.7 \\
    1113  & 1999-11-04 & FAINT & 9.6 \\
    1114  & 1999-11-04 & FAINT & 9.0 \\
    9714  & 2008-03-20 & VFAINT & 29.6 \\
    13993 & 2012-03-21 & VFAINT & 39.6 \\
    13994 & 2012-03-19 & VFAINT & 82.0 \\
    13995 & 2012-03-14 & VFAINT & 63.0 \\
    13996 & 2012-03-27 & VFAINT & 123.1 \\
    14406 & 2012-03-15 & VFAINT & 24.8 \\
    14410 & 2012-03-22 & VFAINT & 78.5 \\
    14411 & 2012-03-20 & VFAINT & 33.6 \\
    14415 & 2012-04-13 & VFAINT & 34.5 \\
    \hline
  \end{tabular}
  \label{tab:chandra}
\end{table}

\subsection{Images}

\begin{figure}
  \centering
  \includegraphics[width=0.9\columnwidth]{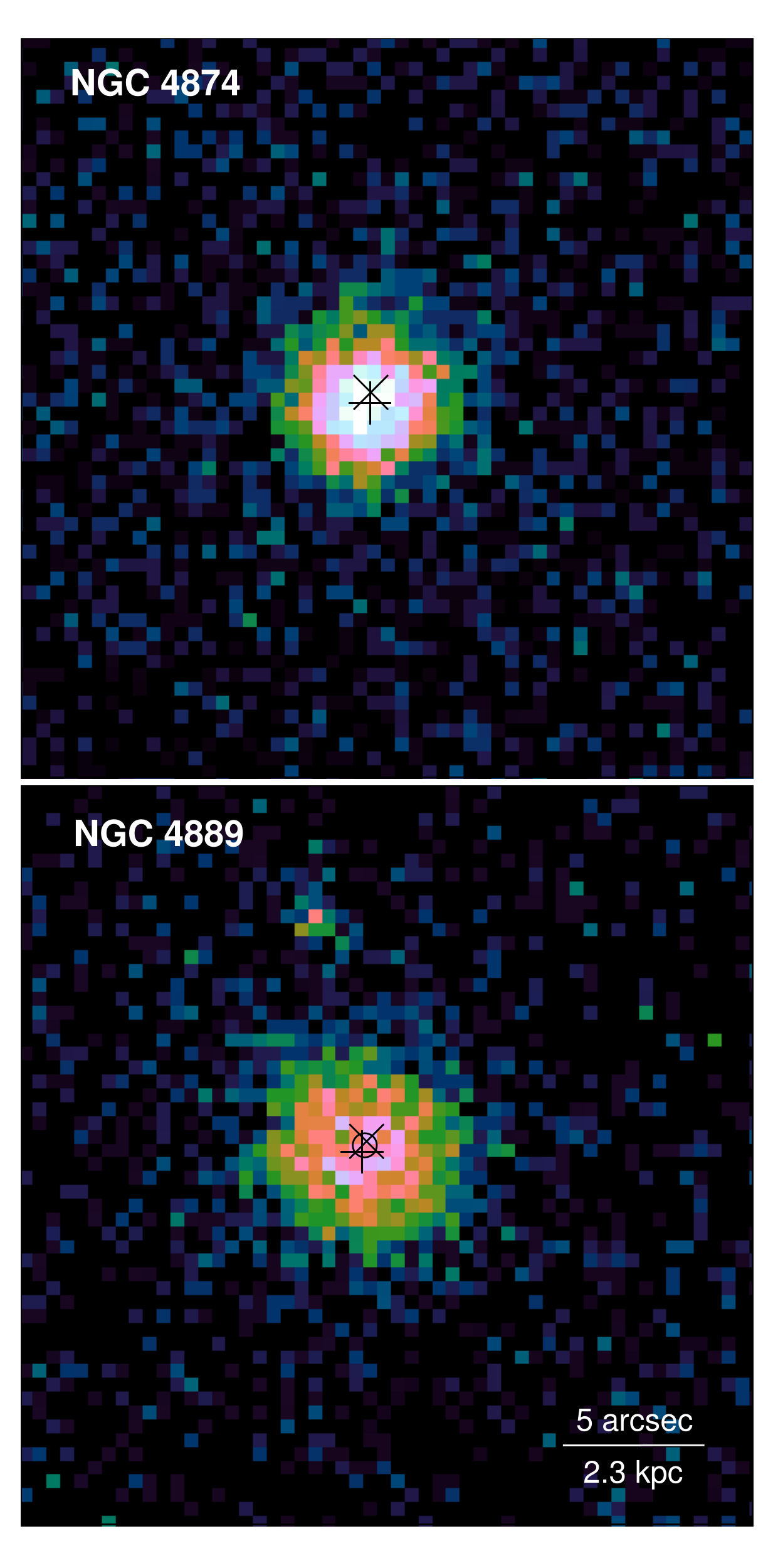}
  \caption{Exposure-corrected X-ray images of the two central galaxies
    in the 0.5 to 3 keV band. We use native 0.492 arcsec pixels. The
    images use the same surface brightness colour scale. The marker
    $\times$ shows the position of the galaxy (NGC\,4874 was taken
    from SDSS; NGC\,4889 was taken from the 1.4~GHz radio source
    position in \protect\citealt{MillerComaRadio09}). The marker $+$
    shows the X-ray centroid for the inner 4.5 arcsec radius. The
    centroids of NGC\,4874 and NGC\,4889 are
    $12^\mathrm{h}59^\mathrm{m}35\fs71$, $+27^{\circ}57'33\farcs0$ and
    $13^\mathrm{h}00^\mathrm{m}08\fs14$, $+27^{\circ}58'36\farcs8$,
    respectively. Marked by $\circ$, the 4.9 GHz radio source in NGC
    4889 is located at $13^\mathrm{h}00^\mathrm{m}08\fs14$,
    $+27^{\circ}58'37\farcs1$ in the VLA image archive. The
      positions are consistent within their uncertainties.}
  \label{fig:img}
\end{figure}

In Fig.~\ref{fig:img} we present X-ray images of the two central
galaxies, showing the exposure-corrected surface brightness in the 0.5
to 3 keV band (chosen to reduce the projected cluster emission). For
the exposure map creation, we assumed a spectral model appropriate for
the cluster emission (9.7~keV temperature, $0.24 \Zsun$ metallicity)
rather than that for the galactic emission, as the cluster background
emission covers a larger spatial region than the galactic
emission. The inner peak of X-ray emission is contained within the
inner 3.5 arcsec radius for both galaxies.

\begin{figure}
  \includegraphics[width=\columnwidth]{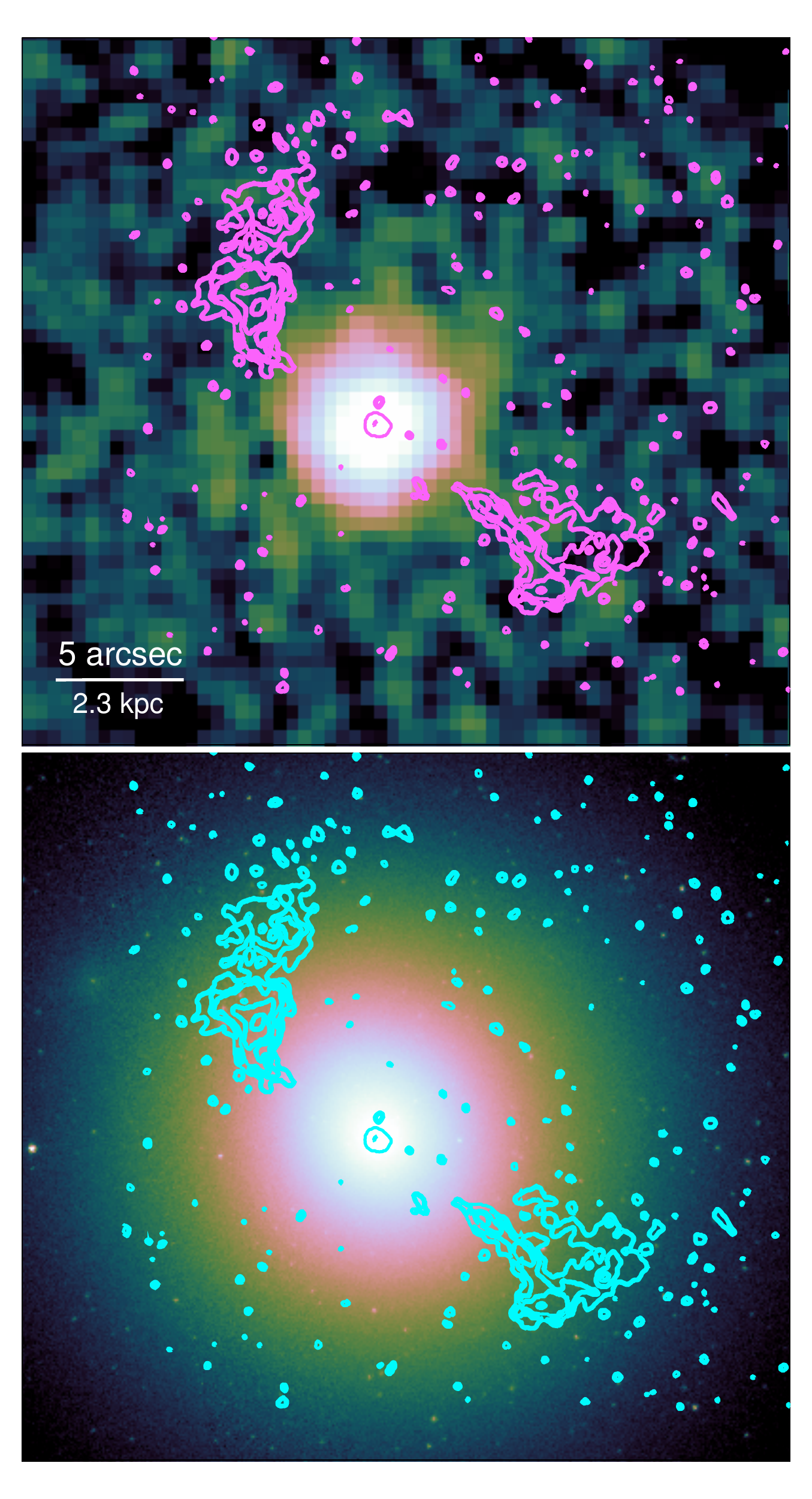}
  \caption{(Top panel) X-ray image of NGC 4874 in the 0.5 to 3 keV
    band, smoothed with a Gaussian of size 2 pixels. The contours are
    6~cm radio data taken from \protect\cite{Feretti87}. (Bottom panel)
    \emph{HST} F814W ACS image of the galaxy with the same contours.}
  \label{fig:radio}
\end{figure}

NGC\,4874 hosts a well known wide-angle-tail (WAT) radio source
\citep{ODeaOwen85}. Fig.~\ref{fig:radio} shows an overlay of the radio
structure on the X-ray and optical images. The jets of the source
appear to be only visible outside the region where the bulk of the
X-ray emission originates in the corona. This anti-correlation was
first pointed out in NGC 4874 by \cite{Sun05_A1367} using lower
resolution radio data. It is outside the corona that the jets will
experience any transverse force due to the relative motion of the
galaxy and ICM.

\begin{figure}
  \includegraphics[width=\columnwidth]{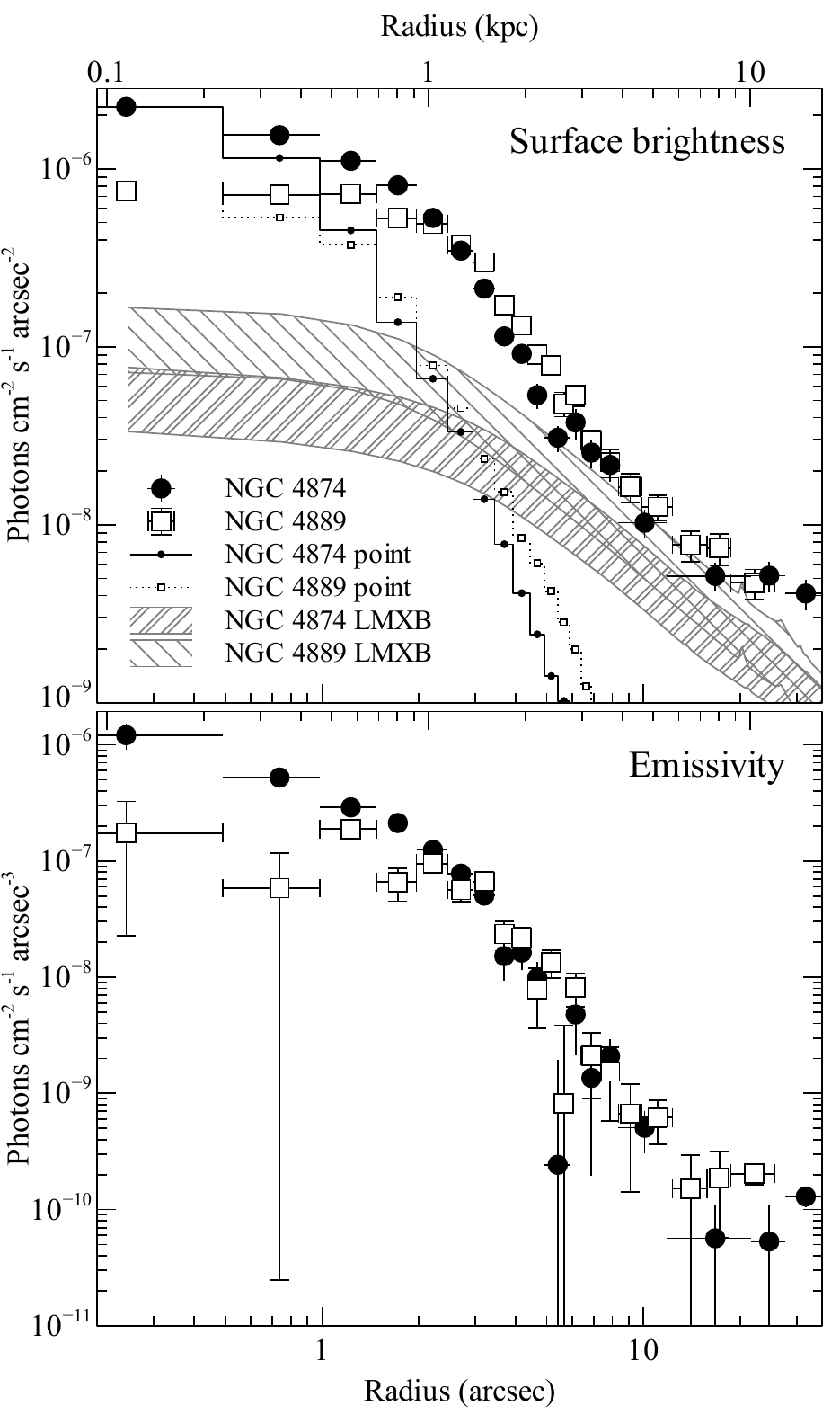}
  \caption{(Top panel) Background-subtracted X-ray surface brightness
    profile of the two galaxies in the 0.5 to 3 keV band. The two
    lines labelled `point' are surface brightness profiles for two
    point sources at the location of the galaxies. We use the X-ray
    spectrum of the respective galaxy and normalise the profiles to
    the central values of their respective galaxies. The regions
    labelled `LMXB' show the expected range of flux from X-ray
    binaries in the galaxies. (Bottom panel) X-ray emissivity
    profiles, created by deprojecting the surface brightness
    profiles.}
  \label{fig:sbprof}
\end{figure}

\begin{figure*}
  \includegraphics[width=\columnwidth]{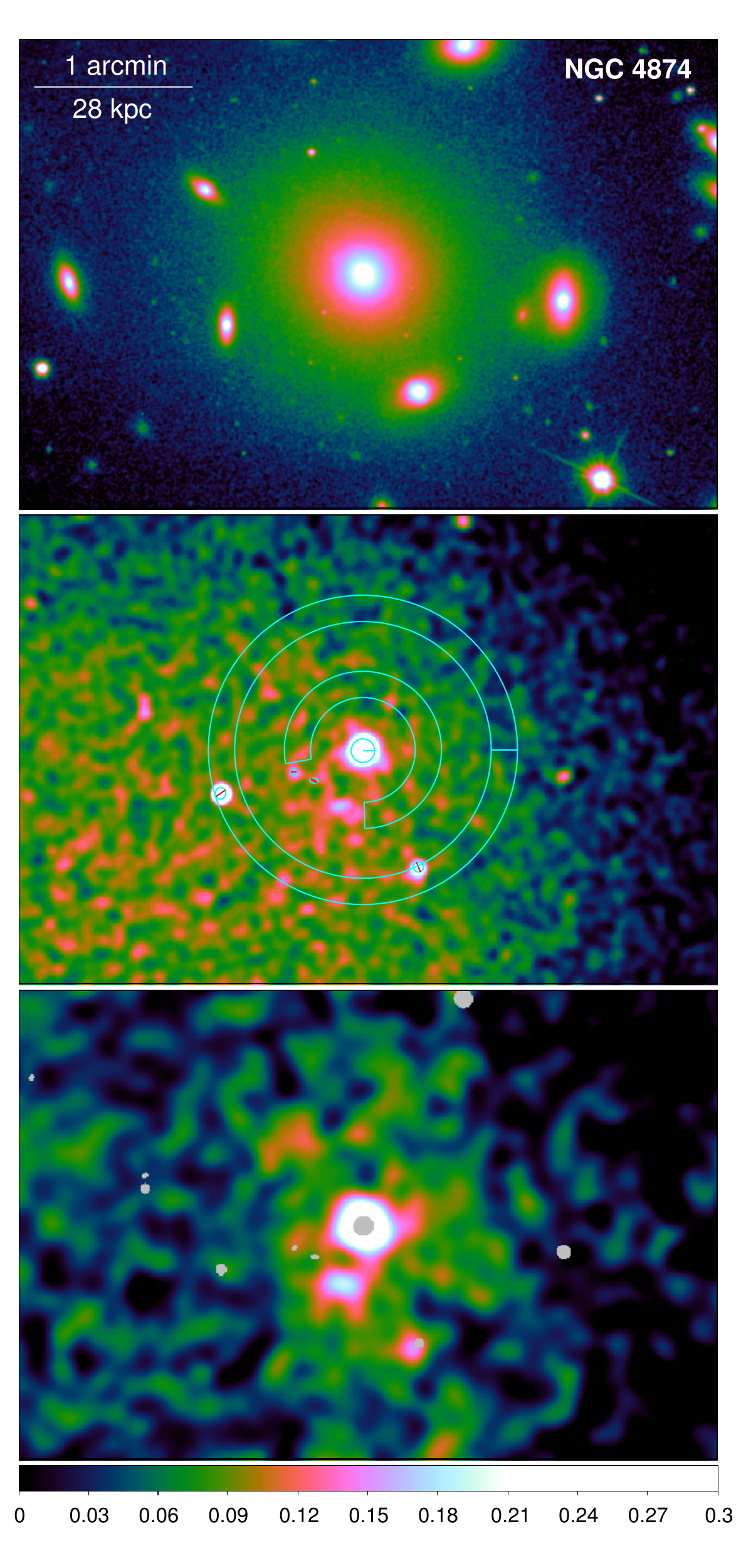}
  \includegraphics[width=\columnwidth]{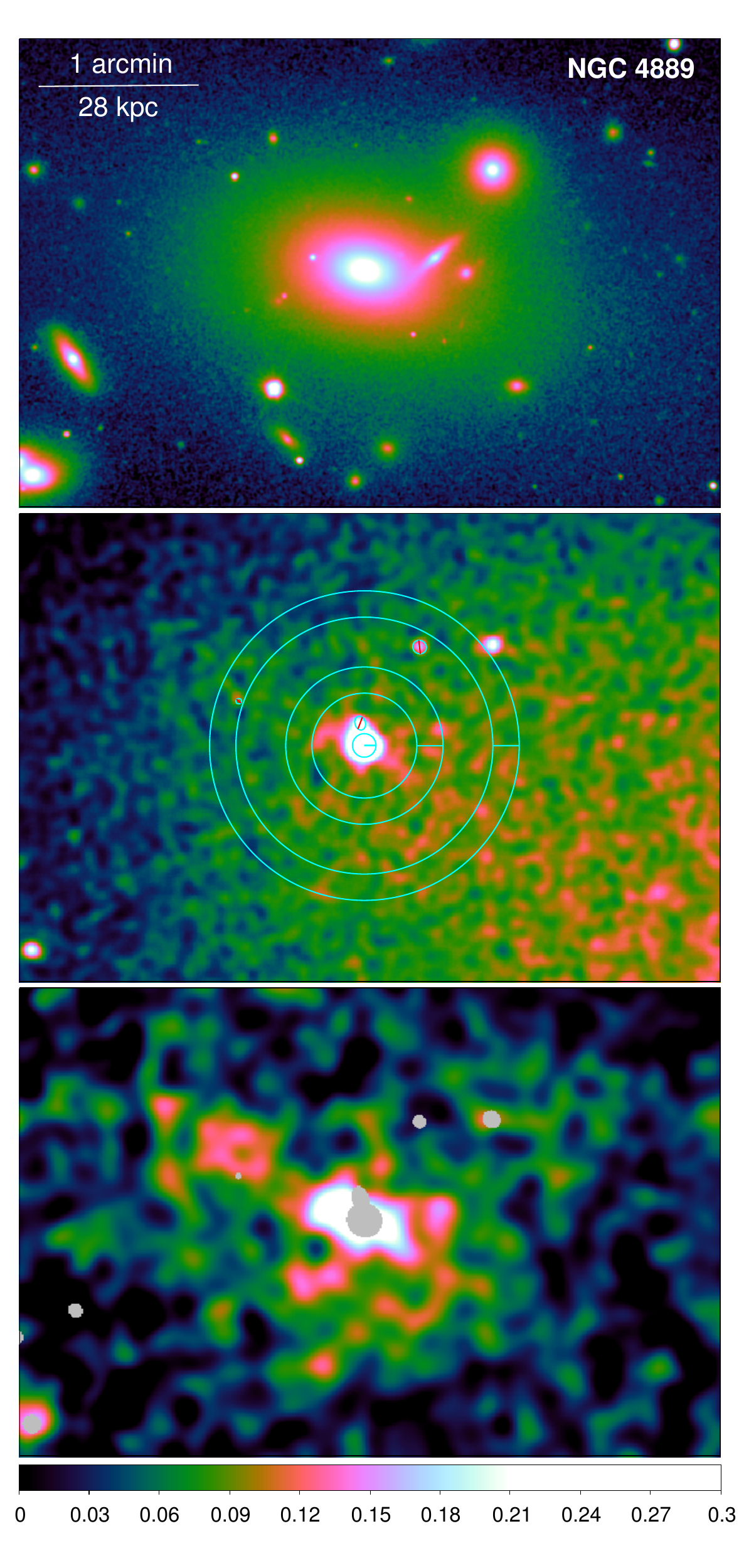}
  \caption{Optical and X-ray images of the outer parts of NGC 4874
    (left) and NGC 4889 (right). (Top row) SDSS r band image. (Centre
    row) 0.5 to 7 keV exposure-corrected-X-ray image of the same
    region, smooth by a Gaussian with $\sigma=1.968$ arcsec. Shown are
    the surface-brightness-background region (outer annulus), spectral
    background region (inner annulus), galaxy spectral region (circle)
    and excluded point source (slashed ellipses). (Bottom row)
    Unsharp-masked image of the same region with point sources, taken
    from \protect\cite{SandersComa13}, excluded (in grey). This shows
    the fractional difference between the X-ray image smoothed by a
    Gaussian with $\sigma=3.94$ and $63.0$ arcsec, with the
      numerical value shown in the colour bars.}
  \label{fig:opt_unsharp}
\end{figure*}

\subsection{Surface brightness profiles}
In Fig.~\ref{fig:sbprof} (top panel) we show surface brightness
profiles of the two galaxies. The X-ray centroid was used as the
centre of the profiles. The profiles were extracted using annuli which
were 0.492 arcsec wide and then were rebinned in radius to have a
minimum signal to noise ratio of 5. Exposure-corrected background
count rates were measured from 49 to 59 arcsec radius
(Fig.~\ref{fig:opt_unsharp}) and subtracted. We excluded point sources
identified in \cite{SandersComa13}. To show the effect of the PSF, we
include on the plot profiles of simulated point sources at the
positions of the galaxies. We simulated the point sources for each
separate observation using \textsc{chart} and \textsc{marx} 5.0.0
\citep{Carter03}, assuming that they have the same spectrum as their
respective galaxy, but ten times the flux. The profiles shown are the
combined set of simulations, each reprojected to the 13994 simulation,
in the 0.5 to 3 keV energy band.

Both of the central galaxies are more extended than expected for point
sources at their locations.  Although the PSF at the location of
NGC\,4889 is larger, this does not account for the noticeably flat
X-ray surface brightness profile in its centre, also seen in
Fig.~\ref{fig:img}. The outer parts of the galaxy have a very similar
surface brightness profile. The centres of the X-ray emission are
consistent within around 0.5 arcsec with the positions of the
galaxies.

The outer X-ray surface brightness profiles are likely to have a
significant component from low-mass X-ray binaries (LMXB). This can be
modelled by taking the optical profile and assuming a constant ratio
between the optical luminosity of the galaxy and its LMXB X-ray
emission. \cite{Sarazin01} examined NGC\,4697, finding a ratio between
the LMXB X-ray luminosity in the 0.3 to 10 keV band and B band optical
luminosity of $8.1 \times 10^{29} \ergps
L_{B\odot}^{-1}$. \cite{Kim09} examined three old elliptical galaxies,
finding a factor of two variation in the ratio between the 0.3 to 8
keV LMXB X-ray luminosity and the K band luminosity of $1-2 \times
10^{29} \ergps L_{K\odot}^{-1}$, apparently increasing with the
globular cluster specific frequency. We show in Fig.~\ref{fig:sbprof}
the expected surface brightness profiles for the LMXB X-ray
contribution. This was calculated by taking the magnitudes of the
galaxies (from \citealt{Sun07}) and converting to a total X-ray count
rates using the different ratios. We assumed a powerlaw LMXB spectrum
with $\Gamma=1.56$ \citep{Irwin03}. \emph{Hubble Space Telescope}
surface brightness profile were made for each galaxy (using filter
F814W) and scaled to create X-ray profiles, using the integrated
optical light (calculated by fitting a S\'ersic profile to the outer
parts and adding the inner contribution). Fig.~\ref{fig:sbprof} shows
the spread in X-ray surface brightness from the different ratios. LMXB
are likely to be important in the outskirts of the galaxies, but the
exact contribution is difficult to estimate. In addition, we have
removed point sources from the data which may be part of this LMXB
contribution.

However, the shape of the measured surface brightness profiles in the
outskirts is also dependent on the region used to extract a background
surface brightness. Fig.~\ref{fig:opt_unsharp} (bottom row) shows that
the faint X-ray emission around each galaxy has a complex
morphology. It is not obvious what part of that emission is associated
with the galaxy and what part with the cluster. If this extended
emission has a strong LMXB contribution, then the LMXB emission is not
strongly correlated with the optical morphology of the galaxies.

If the WAT source shape indicates the path of NGC\,4874 through the
cluster, it is heading in a south-east direction
\citep{Feretti87}. Any material stripped from the corona should
appear behind the path of the galaxy. We do not see any evidence for
excess X-ray emission (Figs. \ref{fig:radio} and
\ref{fig:opt_unsharp}) from such stripped material.

Fig.~\ref{fig:sbprof} (bottom panel) shows the 3D radial emissivity,
which is deprojected from the surface brightness under the assumption
of spherical symmetry. The emissivity error bars are generated by
repeating the deprojection with different input profiles, by making
different realisations based on the mean value and the size of the
error bars, and examining the distribution of output profiles. For
NGC\,4889 the emissivity within a radius of 0.2~kpc suddenly
drops. The region may be surrounded by a rim of high emissivity (the
third innermost radial bin).

\subsection{Image fitting}
\begin{figure}
  \includegraphics[width=\columnwidth]{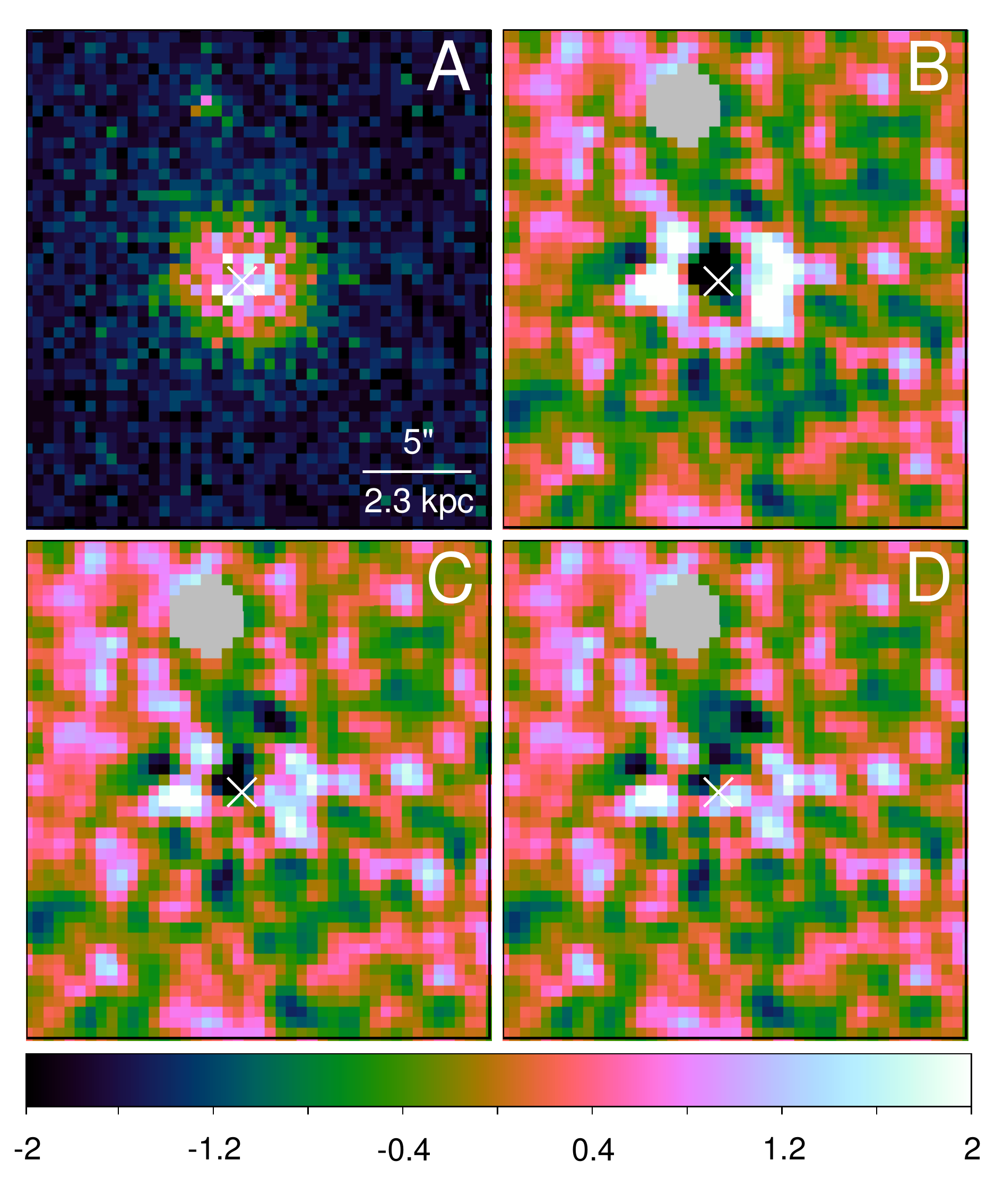}
  \caption{Residuals of surface brightness fits to the NGC 4889
    region, accounting for PSF. The point source to the north (grey)
    is excluded. (A) Original X-ray image (using a different colour
    scaling), (B) residuals from a $\beta$ model with fixed
    $\beta=2/3$, (C) residuals from a $\beta$ model with free $\beta$
    parameter and (D) residuals from a $\beta$ model with a central
    cavity. The residual images have been smoothed by a
      Gaussian with $\sigma=1$ pixel (0.492 arcsec). The colour bar
      shows the numerical value in counts per pixel.}
  \label{fig:ngc4889residuals}
\end{figure}

We can examine this hole in more detail including the effect of the
PSF. We used \textsc{sherpa} \citep{Freeman01Sherpa} to fit the X-ray
count image in a $64\times 64$ pixel box around NGC 4889. We
implemented a density model where the density was zero inside
$r_\mathrm{cavity}$ and a standard $\beta$ model outside that
to some maximum radius ($r_\mathrm{max}$). This was then
projected onto the sky to make a surface brightness image. The model
was fitted to the data by minimising the C-statistic, after
convolution by the PSF generated using the \textsc{chart} and
\textsc{marx} simulations. Fig.~\ref{fig:ngc4889residuals} shows the
residuals from the data for three variations of the model, (B)
  fixing $\beta = 2/3$ with $r_\mathrm{cavity}=0$, (C) a free $\beta$
  parameter with $r_\mathrm{cavity}=0$ and (D) allowing $\beta$ and
$r_\mathrm{cavity}$ to vary. The final model (D) improved the fit
statistic by $10.9$ over model (C). Model (D) implies a
$r_\mathrm{cavity}=0.6$~kpc.

If the corona of the galaxy is not spherical in shape, it is possible
that we could erroneously infer the presence of a cavity. For this to
be the case, the corona would have to be a disc lying in the plane of
the sky. Any projected emission in front or behind the core should
increase the surface brightness.  Its morphology should be spherical,
unless it is significantly affected by ram pressure.

\begin{table}
  \caption{Parameters from $\beta$ model fits to images.}
  \centering
  \begin{tabular}{llccc}
    \hline
    Object & Parameter & Fixed $\beta$ & Variable $\beta$    & Inner cavity \\
    \hline
    NGC 4889 & $\beta$   & $2/3$         & $1.5_{-0.3}^{+0.5}$ & $1.1_{-0.2}^{+0.1}$ \\
             &$r_\mathrm{max}$ (arcsec) & $>19$ & $>10$ & $>10$ \\
             &$r_\mathrm{cavity}$ (arcsec) & - & - & $1.2_{-0.1}^{+0.2}$ \\
    \hline
    NGC 4874 &  $\beta$ & $2/3$ & $0.80 \pm 0.06$ & $0.79 \pm 0.05$ \\
             &$r_\mathrm{max}$ (arcsec) & $>13$ & $>12$ & $>13$ \\
             &$r_\mathrm{cavity}$ (arcsec) & - & - & $<0.5$ \\
    \hline
  \end{tabular}
\end{table}

\begin{figure}
  \includegraphics[width=\columnwidth]{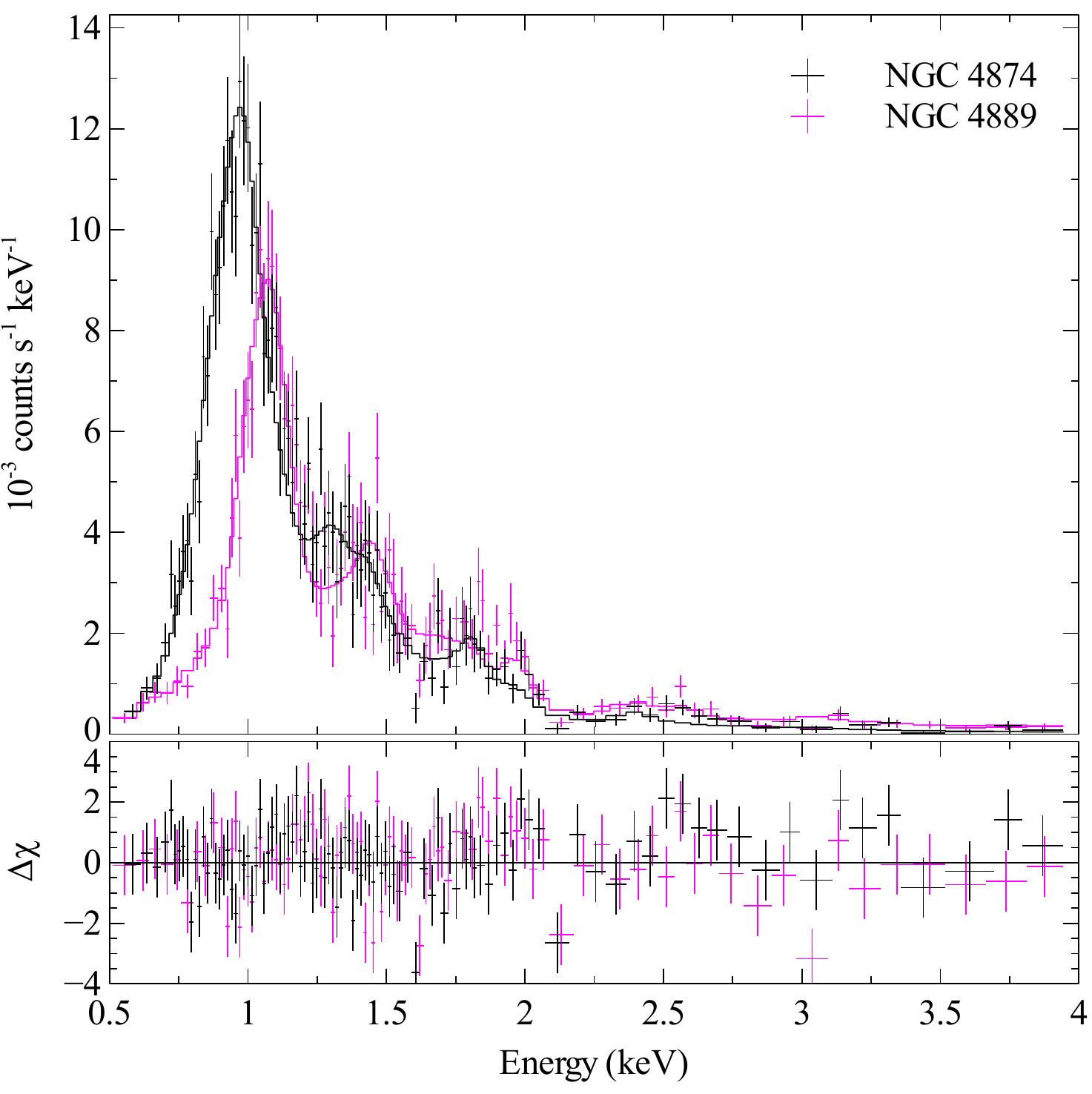}
  \caption{Spectral fit to the background-subtracted X-ray spectra of
    the galaxies. In these fits the Galactic column density was
    allowed to be free. Source spectra were extracted from regions 4.5
    arcsec in radius. Background spectra were extracted from an
    annular region 20 to 30 arcsec radius. For NGC 4874 an angular
    range starting 10 degrees south from the eastern direction and
    ending in the south direction, was excluded to avoid some point
    sources (Fig.~\ref{fig:opt_unsharp}).}
  \label{fig:specfit}
\end{figure}

\begin{table*}
  \centering
  \caption{Spectral fit parameters for the two galaxies. We show the
    parameters allowing the Galactic column density to be free in the
    fit, or fixing it at the value of $0.85 \times 10^{20} \pcmsq$
    from \protect\cite{Kalberla05}. The
    data were fit between 0.5 and 4 keV, minimising the $\chi^2$
    statistic.  We also show the results where the spectra were jointly
    fit with the background spectra, fitting the data between 0.5 and 7
    keV (1T). In this case an \textsc{apec} model was fit to the corona
    and background simultaneously, scaling by the areas on the sky and
    with an additional component for the coronal emission.  We list the
    temperature and metallicity of the background component as
    ``Background''.  For NGC 4874, we also use a model with a second
    coronal temperature component, with the metallicity tied to the
    first component (2T). The flux shown is between 0.5
    and 4 keV. The luminosity is bolometric, with absorption
    removed. The number of counts is between 0.5 and 4 keV after
    subtracting the projected cluster emission (around 20 per cent of
    the counts).  }

  \begin{tabular}{llllll}
    \hline
    Object   & Parameter                        & Fit $n_\mathrm{H}$     & Fix $n_\mathrm{H}$    & Joint fit (1T)     & Joint fit (2T) \\
    \hline
    NGC 4874 & $n_\mathrm{H}$ ($10^{20}\pcmsq$) & $15_{-4}^{+5}$         & $0.85$                & $4.4 \pm 0.6$      & $4.5 \pm 0.7$ \\
             & $kT_1$ (keV)                     & $1.04 \pm 0.02$        & $1.18 \pm 0.02$       & $1.14 \pm 0.3$     & $1.04 \pm 0.02$ \\
             & $kT_2$ (keV)                     &                        &                       &                    & $2.2 \pm 0.3$ \\
             & $Z$ ($\Zsun$)                    & $0.40_{-0.06}^{+0.09}$ & $0.40 \pm 0.06$       & $0.37 \pm 0.06$    & $1.3_{-0.4}^{+0.6}$ \\
             & norm$_1$ ($10^{-5}$ cm$^{-5}$)   & $5.3 \pm 0.5$          & $4.2 \pm 0.4$         & $4.6 \pm 0.4$      & $1.1_{-0.3}^{+0.4}$ \\
             & norm$_2$ ($10^{-5}$ cm$^{-5}$)   &                        &                       &                    & $1.2 \pm 0.2$ \\
             & Background $kT$ (keV)            &                        &                       & $7.7 \pm 0.4$      & $7.5 \pm 0.4$ \\
             & Background $Z$ ($\Zsun$)         &                        &                       & $0.23 \pm 0.06$    & $0.22 \pm 0.06$ \\
             & $\chi^2_\nu$                     & $117 / 98 = 1.19$      & $133 / 99 = 1.34$     & $492 / 426 = 1.15$ & $470 / 424 = 1.11$ \\
             & Flux ($10^{-14} \ergpcmsqps)$    & $4.2$                  & $4.4$                 & $4.3$              & $4.4$ \\
             & Luminosity ($10^{40}\ergps$)     & $12$                   & $9.0$                 & $9.9$              & $9.1$ \\
             & Counts                           & $3335$ \\
    \hline
    NGC 4889 & $n_\mathrm{H}$ ($10^{20}\pcmsq$) & $9 \pm 3$              & $0.85$                & $6.3 \pm 0.7$      \\
             & $kT$ (keV)                       & $1.9 \pm 0.1$          & $2.1 \pm 0.4$         & $2.0 \pm 0.06$     \\
             & $Z$ ($\Zsun$)                    & $1.5 \pm 0.4$          & $2.2^{+0.5}_{-0.4}$   & $1.7 \pm 0.3$      \\
             & norm ($10^{-5}$ cm$^{-5}$)       & $3.0 \pm 0.5$          & $2.2 \pm 0.3$         & $2.8 \pm 0.3$      \\
             & Background $kT$ (keV)            &                        &                       & $6.8 \pm 0.3$      \\
             & Background $Z$ ($\Zsun$)         &                        &                       & $0.28 \pm 0.06$    \\
             & $\chi^2_\nu$                     & $123 / 93 = 1.32$      & $136 / 94 = 1.45$     & $467 / 427 = 1.10$ \\
             & Flux ($10^{-14} \ergpcmsqps)$    & $3.9$                  & $4.0$                 & $4.0$              \\
             & Luminosity ($10^{40}\ergps$)     & $9.1$                  & $7.9$                 & $8.8$              \\
             & Counts                           & $3768$ \\
    \hline
  \end{tabular}
  \label{tab:params}
\end{table*}

\subsection{Spectral fitting}
\label{sect:fitting}
Fig.~\ref{fig:specfit} shows fits to the spectra extracted from the
two galaxies, after subtraction of the background cluster
emission. They were fit between 0.5 and 4 keV using an \textsc{apec}
emission model \citep{SmithApec01}, version 2.0.1, with \textsc{phabs}
photoelectric absorption to account for Galactic absorption
\citep{BalucinskaChurchPhabs92}. To analyse the data we used
\textsc{ciao} version 4.5 and \textsc{caldb} version 4.5.9, which
includes recent corrections for the optical blocking filter (OBF)
contamination. When creating response matrices and ancillary response
matrices, we weighted spatial regions by the number of counts in the 0.5 to
7 keV band. The spectra and background spectra were added
together. The response matrices and ancillary response matrices were
weighted by the relative exposure times.

\begin{figure}
  \includegraphics[width=\columnwidth]{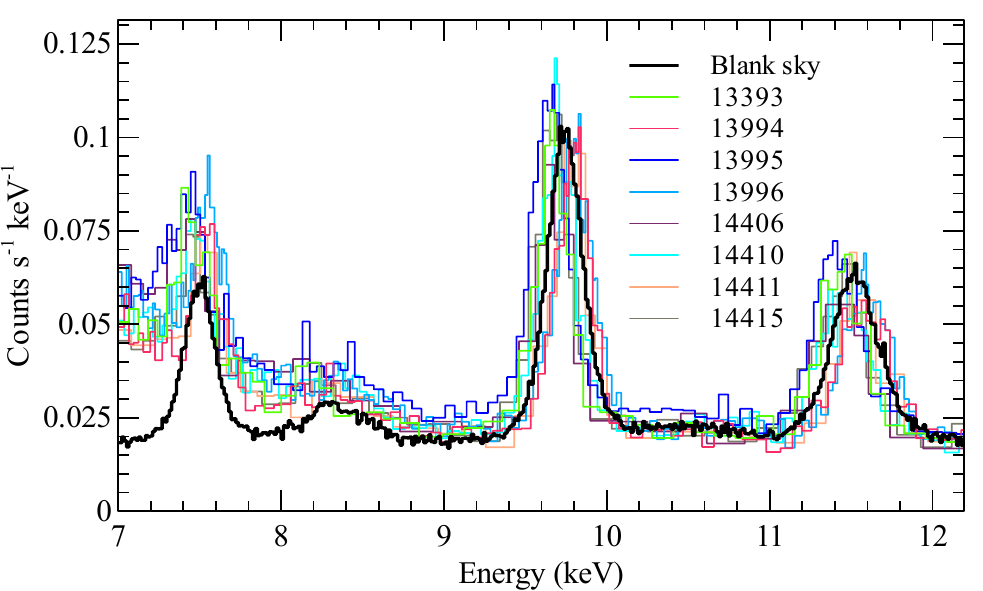}
  \caption{Comparison of the spectra of the instrumental lines in the
    observations taken in 2012 for the ACIS-I2 CCD. Also plotted is a
    spectrum taken from blank sky observations, appropriate for the
    time period. The spectra have been rebinned to have a signal to
    noise of 10 per spectral bin.}
  \label{fig:cal_compare}
\end{figure}

\begin{figure}
  \includegraphics[width=\columnwidth]{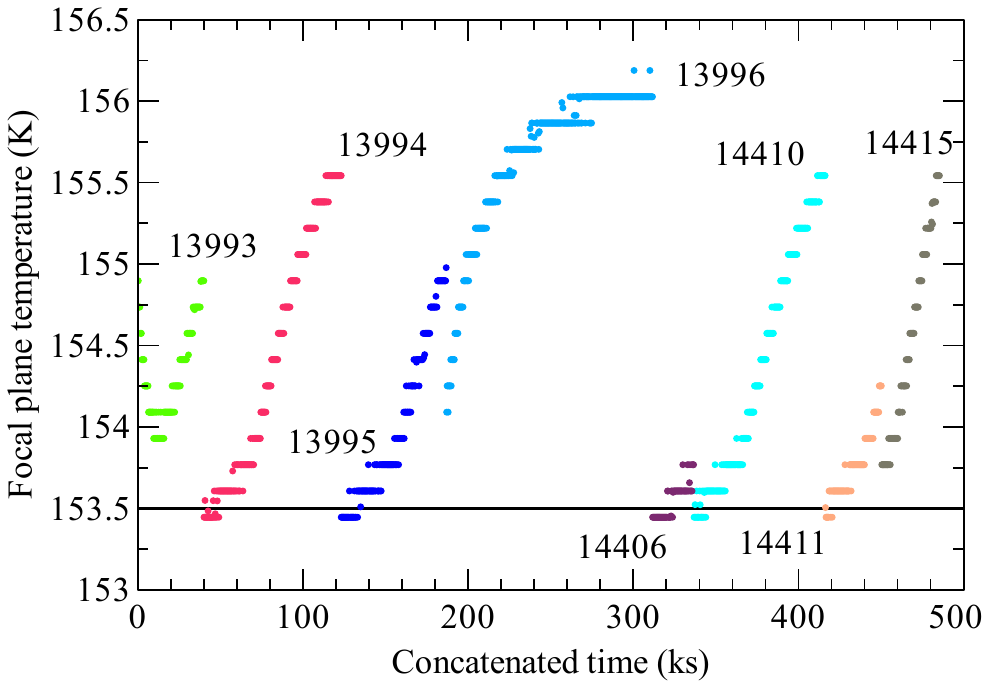}
  \caption{Focal plane temperature during the 2012 observations. The
    observation times are relative to the end of the time of previous
    observation identifier.  The horizontal line shows the ACIS
    nominal temperature.}
  \label{fig:fp_temp}
\end{figure}

The parameters obtained from the spectral fits are shown in Table
\ref{tab:params}. The first set of spectral fits allow the Galactic
column density ($n_\mathrm{H}$) to be free. The values from the fits
are significantly higher than the value inferred from H\textsc{i}
measurements ($0.85\times 10^{20}\psqcm$; \citealt{Kalberla05}). The
best fitting $n_\mathrm{H}$ values are also unaffected by changing
between the contaminant model in \textsc{caldb} version 4.5.9 and
previous versions. The likely explanation is that there are
significant problems with the instrumental gain calibration during
these Coma observations. If event energies are artificially increased,
this can appear to be an increase in absorbing column
density. Fig.~\ref{fig:cal_compare} compares the spectra of the
instrumental lines at high energy in the observations taken in 2012,
showing that the energy calibration at these energies is not
stable. The cause for this gain variation appears to be large changes
in the focal plane temperature during the observations
(Fig.~\ref{fig:fp_temp}). In particular, observation 13996 spends long
periods away from the nominal temperature at which the instrument is
calibrated.

Unfortunately, these gain changes cannot easily be compensated for in
the analysis. We attempted to find the gain variation by jointly
fitting the spectra of the surrounding cluster emission (0.4 to 1.4
arcmin radius) from each of the observations. The gain was allowed to
vary using the gain fit feature in \textsc{xspec} with the column
density fixed to the H\textsc{i} value.  If these gain corrections
were then applied to the spectra of the galaxies, they did not appear
to significantly reduce the best fitting column density. This lack of
improvement could be because the gain variations are not simple linear
shifts. In addition, residual spectral features not adequately fit by
the spectral model may affect the best fitting gain values.

We therefore caution that the spectral parameters we derive
are subject to systematic uncertainties. When fitting for gain
variations and applying them to the combined spectra of the corona, we
obtained for NGC 4889 changes in temperature of around 0.2 keV,
abundance $0.2 \Zsun$ and 20 per cent normalisation. The likely
systematic uncertainties from the gain calibration are likely of this
order. If the spectra from each observation were jointly fit for
NGC\,4889, allowing for different gain shifts in each, smaller changes
than this were found.

As the surrounding background regions are similarly affected by gain
issues, to investigate the systematics further, we jointly fit the
background cluster spectra (using the regions shown in
Fig.~\ref{fig:opt_unsharp}) and the spectra of coronae.  The fits used
a common \textsc{apec} component to account for the cluster emission
and another component for the coronal emission. Both components were
absorbed by a common absorption model. The resulting spectral
parameters (Table \ref{tab:params}), except for $n_\mathrm{H}$, are
similar to the previous fits where the background was subtracted from
the coronal spectra. We also investigated the use of earlier versions
of the \textsc{apec} spectral model which has significantly different
models for the Fe-L spectral lines. If version 1.3.1 is used instead
of 2.0.1, the best fitting temperatures change slightly to 0.99 and
1.79 keV for NGC 4874 and 4889, respectively, when fitting for
$n_\mathrm{H}$.

The spectrum of NGC\,4874 allows the inclusion of a second thermal
component with a temperature of around 2~keV. The addition of this
component significantly reduces the $\chi^2$ of the fit and raises the
best fitting metallicity to $1.3 \Zsun$ (assuming both coronal
components have the same metallicity). The hotter second component is
likely due to material at larger radius in the corona. If a smaller
extraction and background region are used, the hotter component
reduces in significance. A second component added to the NGC\,4889
spectral fit cannot have its temperature constrained.

\begin{figure}
  \includegraphics[width=\columnwidth]{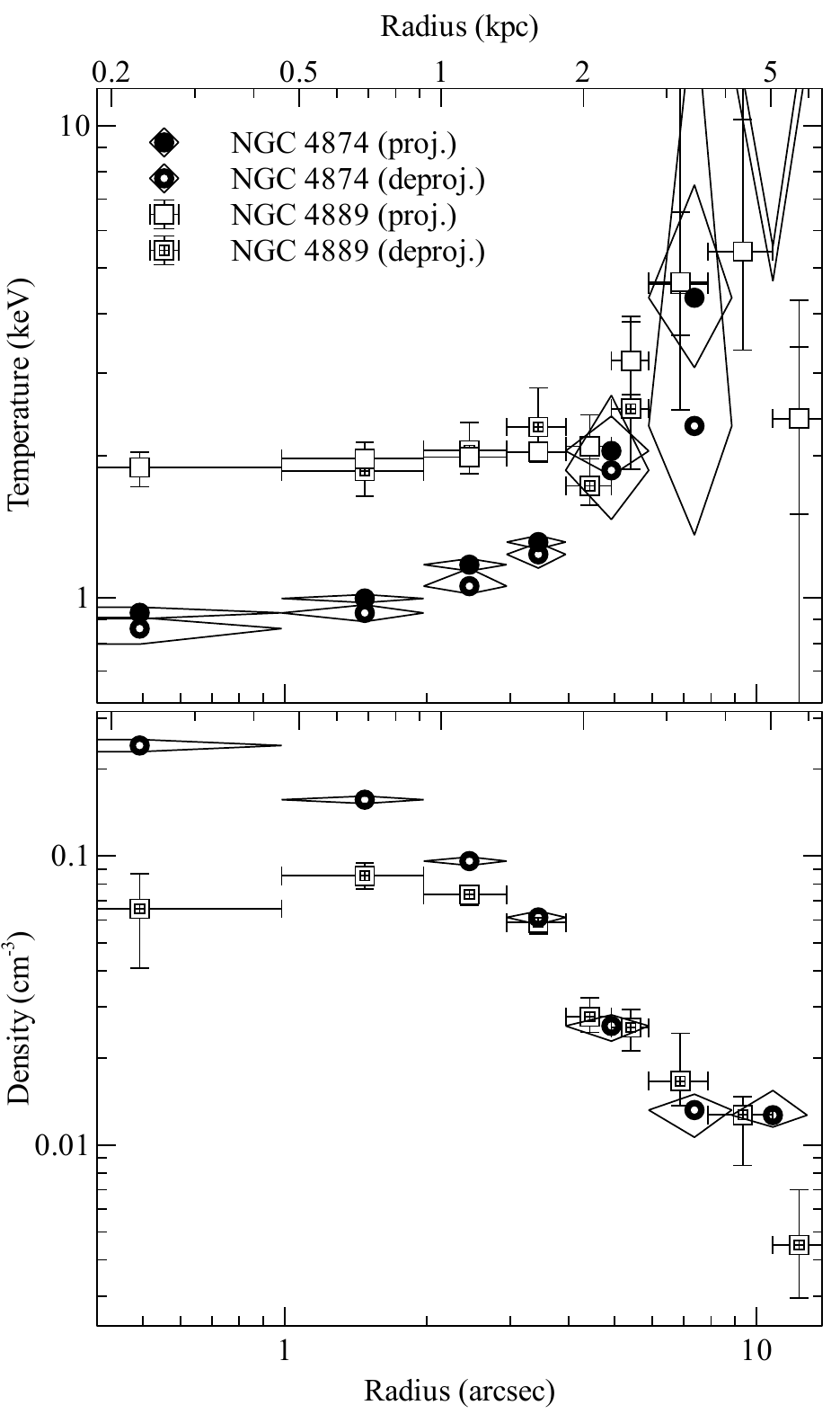}
  \caption{(Top panel) Temperature profiles for the galaxies, fitting
    projected and deprojected spectra. (Bottom panel) Electron density
    profiles, calculated from the normalisations of the deprojected
    spectral fits. Note that these fits do not include the spectral
    contribution from LMXB emission. The deprojected temperature of
    the innermost annulus of NGC\,4889 could not be constrained, so
    the density is calculated assuming the projected temperature.}
  \label{fig:profiles}
\end{figure}

To examine the properties of the coronae as a function of radius we
examined spectra extracted from annular regions.  In
Fig.~\ref{fig:profiles} we plot the spectral profiles of the density
and temperature for the two galaxies. The temperatures shown were
obtained by fitting the projected spectra and deprojected spectra
calculated using the \textsc{dsdeproj} algorithm
\citep{SandersPer07,Russell08}.  The spectra were fit with an
\textsc{apec} thermal model, fixing the metallicity and absorbing
Hydrogen column density to the best fitting single component value
(Table \ref{tab:params}). The profiles show that the effect of
projection in these spectra is relatively small. As was found by
\cite{Vikhlinin01}, the temperature in NGC\,4874 rises with radius in
the inner few arcsec. In NGC\,4889, there may be a mild rise in
temperature with radius. The temperatures rapidly rise at a radius of
2.5 to 5 kpc, where the surface brightness is rapidly declining.  We
note, however, that in both galaxies there may be a significant
contribution to the spectra from LMXB emission in the outer parts. The
density profile for NGC\,4889, like the emissivity, shows a depression
in density in the central region.

We looked for non-thermal powerlaw emission from the nuclei of the two
galaxies. We examined the emission inside 1.5 arcsec radius from the
X-ray centroid for the two galaxies, using the a region of 1.5 to 2.5
arcsec as a background region. A model comprising of an \textsc{apec}
component plus a powerlaw with its photon index fixed to 1.7 was
fitted to the spectra. We used a range of fixed values for the
absorption and metallicity, taken from Table \ref{tab:params}. We
obtained upper limits for 2--10 keV X-ray luminosity for NGC 4874 and
4889 of $1 \times 10^{39}$ and $3\times 10^{39} \ergps$, respectively,
at the 90 per cent confidence level. Using a smaller extraction region
for NGC 4889 (1 arcsec, with 1 to 2 arcsec background) gives an
improved limit of $1 \times 10^{39} \ergps$.

The metallicity from the single temperature component fit of NGC\,4874
appears to be significantly lower than for NGC\,4889 ($0.37$ compared
to $1.5 \Zsun$, respectively). However, if a second thermal component
is included in the NGC\,4874 spectral fit, its metallicity increases
to be consistent with NGC\,4889.  The measurement of metallicities at
$\sim 1 \keV$ temperatures is difficult due to the weak continuum
relative to the line emission. Examining the residuals of the single
temperature component model to the NGC\,4874 data shows an excess
emission above 4 keV. This is likely due to the temperature gradient
within the corona or it could be due to an increasing contribution of
LMXB powerlaw emission with radius. We note that previously
\cite{Vikhlinin01} obtained a metallicity for NGC\,4874 of
$0.79^{+0.84}_{-0.22}$ using a single component, which is greater than
our single component fit result. It is possible that this difference
may be due to the gain calibration uncertainties. We verified that the
choice of \textsc{mekal} or \textsc{apec} did not significantly change
the results of the spectral fits. The background region used by
\cite{Vikhlinin01} (15 to 30 arcsec radius) is similar to the one used
here (20 to 30 arcsec radius), so this is unlikely to cause the
metallicity difference.

We can also attempt to determine the metallicities of the coronae
relative to Fe to examine enrichment scenarios. However, the results
depend on the spectral code used. For NGC 4874, we fit the
background-subtracted spectrum between 0.5 and 7 keV, forcing the
column density to the value from fitting the cluster material and
allowing the $\alpha$-elements to vary in their solar proportions
relative to iron, with the other elements fixed to have the same ratio
to solar as iron, we obtain a relative ratio of $\alpha$ to Fe of $1.3
\pm 0.2 \, Z_{\odot}/Z_{\odot}$, using the \textsc{apec}
model. However, if Ni is allowed to vary, the peak emission in the
Fe-L region is instead fit by a very high Ni to Fe ratio ($\sim 13$),
changing the $\alpha$ to Fe ratio to 2.3. Using the CIE model from
\textsc{spex} \citep{Kaastra96} gives an $\alpha$ to Fe ratio of $1.1
\pm 0.2$, if Ni is fixed to Fe, or $1.4 \pm 0.2$, if Ni is allowed to
vary (giving a Ni to Fe ratio of $5.0_{-1.4}^{+2.1}$). For NGC\,4889,
we obtain similar results where the $\alpha$ to Fe ratios are $1.2 \pm
0.2$ and $1.0 \pm 0.2$, for \textsc{apec} and \textsc{spex},
respectively, if Ni is tied to Fe. If Ni is free to vary, the ratios
change to $1.9^{+0.6}_{-0.4}$ and $1.1 \pm 0.2$, respectively. We note
that the implemented ionization balance for Ni in the current plasma
codes is inaccurate and can significantly bias the Ni abundance
measurements.  Taking into account this and the gain calibration
uncertainties, the metallicity ratios in the coronae appear consistent
with Solar values.

\subsection{Stripped tail}

\begin{figure}
  \includegraphics[width=\columnwidth]{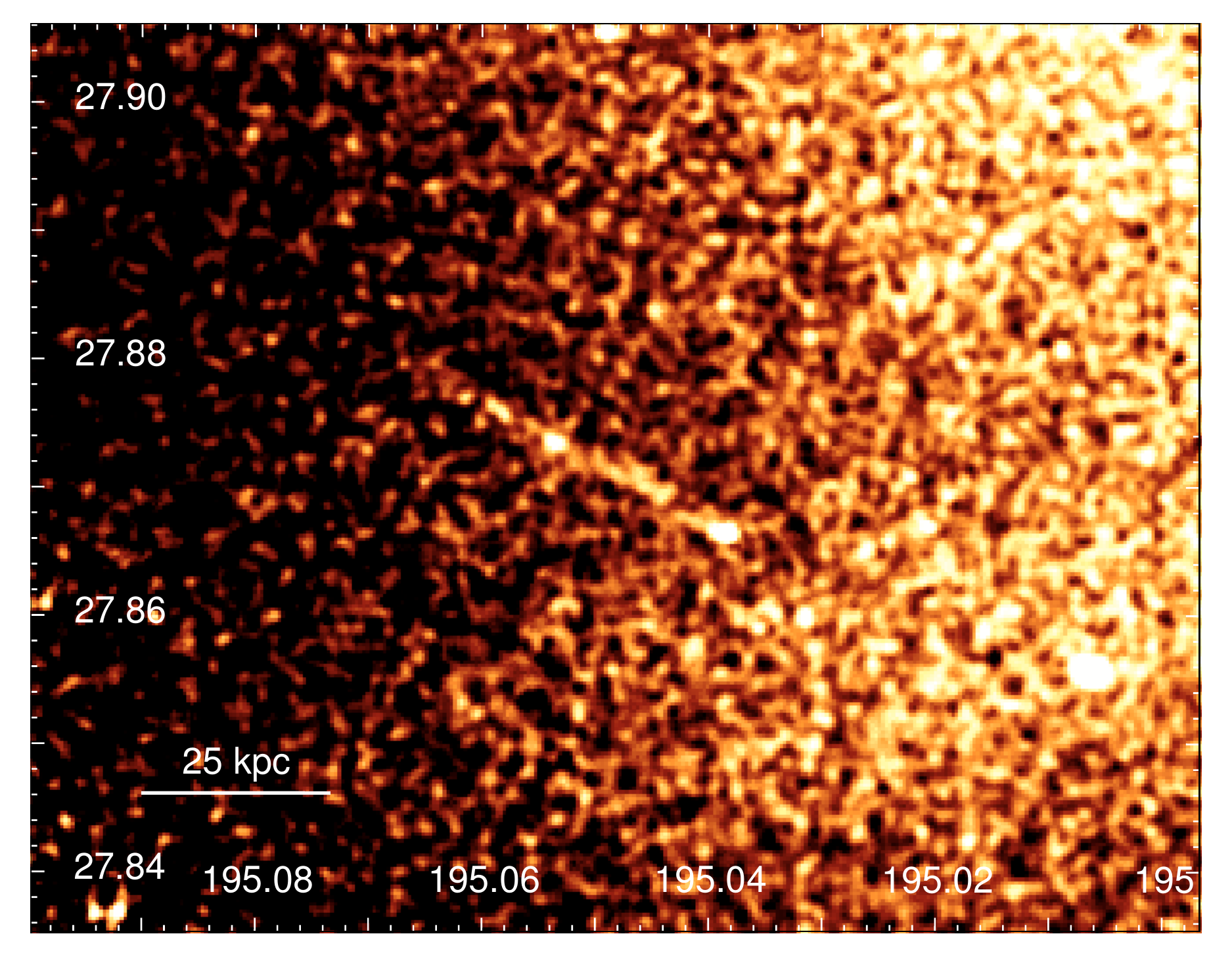}
  \caption{0.5 to 1.5 keV X-ray image of the tail behind galaxy D100 /
    GMP\,2910. The image was binned into 0.984 arcsec pixels and
    smoothed by 3 pixels.}
  \label{fig:d100}
\end{figure}

\cite{Yagi07} showed the existence of a 60-kpc-long filament of
H$\alpha$ emission behind the galaxy GMP\,2910 (also known as
D100). As discussed in \cite{SandersComa13}, there is soft
X-ray emission associated with the filament (Fig.~\ref{fig:d100}). The
extent and location of the tail appears to be the same as the
H$\alpha$ emission (50 kpc), except for the faintest structure at the end
furthest from the galaxy. The filament appears to be $\sim 7$~arcsec
wide or around 3 kpc.

\begin{figure}
  \includegraphics[width=\columnwidth]{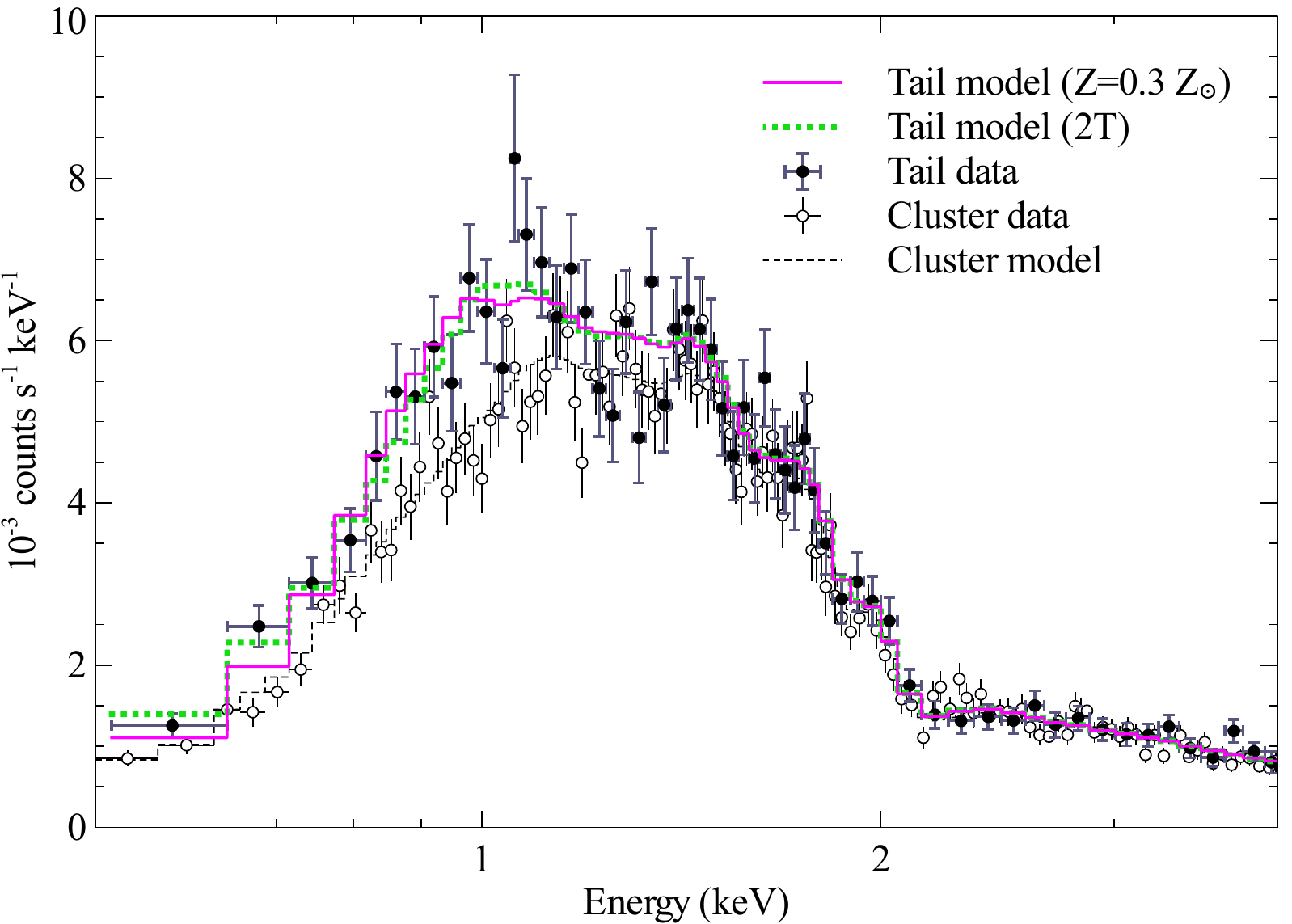}
  \caption{Spectrum in the tail region compared to the nearby cluster
    spectrum. The cluster spectrum has been scaled to the same area as
    the tail and both have been rebinned for display to have a minimum
    signal to noise ratio of 8.  Shown are models for the cluster
    emission and the cluster plus tail model spectrum assuming
    $0.3\Zsun$ metallicity for the tail. We also show a model where a
    second thermal component was fit to the tail (2T).}
  \label{fig:tailspec}
\end{figure}

To examine the spectrum of the tail we fit a joint model to the tail
and the surrounding projected emission (Fig. \ref{fig:tailspec}).  The
spectra were fit simultaneously between 0.5 and 4 keV, minimising the
C-statistic. The emission from the tail was assumed to be the cluster
emission (scaled by area) plus its own thermal component. In the fits
the absorbing column density was free, but linked between the two
components.

Using a single component \textsc{apec} for the tail, its temperature
is $1.0 \pm 0.1$ keV assuming its metallicity is $0.3 \Zsun$. The best
fitting value of $n_\mathrm{H}$ is $(1.5 \pm 0.9) \times 10^{20}
\pcmsq$.  The bolometric luminosity of the tail is $1 \times 10^{40}
\ergps$ and the spectrum has 538 net counts above the cluster
background in the 0.5 to 7 keV band. The 0.5 to 2 keV flux is $6
\times 10^{-15} \ergpcmsqps$. If the metallicity is allowed to vary,
its best fitting value is rather low at $0.09_{-0.06}^{+0.13} \Zsun$,
which gives a lower $0.9 \pm 0.1$ keV temperature.

A second thermal \textsc{apec} component can be included in the
spectral fit for the tail, assuming both components have the same
metallicity. In this case, the two temperatures obtained are
$0.21_{-0.04}^{+0.07}$ and $1.3_{-0.3}^{+0.4}$ keV and $n_\mathrm{H}$
is $(2.3 \pm 0.9) \times 10^{20} \psqcm$. The C-Statistic for the
spectral fit reduces from 317 to 311 (changing from 365 to 362 degrees
of freedom). With the second component, the best fitting metallicity
is $0.6_{-0.5}^{+1.9} \Zsun$. The bolometric luminosity in this case
increases to $4 \times 10^{40} \ergps$ and the flux is $9 \times
10^{-15} \ergpcmsqps$. Similar low temperature single-component
metallicities and second temperature components have been see in
spectra from other stripped tails \citep[e.g.][]{Zhang13}. We examined
the spectra of the front and back halves of the tail, fitting single
thermal components for the tail emission. We found that they have
consistent $\sim 1$~keV temperatures.

Taking the single component model, the electron density of
  the X-ray emitting material in the tail is $8 \times 10^{-3}
  \pcmcu$, assuming the emission fills a cylinder 2 kpc in radius and
  36 kpc in length. If the material is clumpy, this value would be a
  lower limit. The density gives a gas mass of $10^{8} \Msun$, similar
  to the amount of material in ionised gas \citep{Yagi07}.  If the
  metallicity of the material is $0.3 \Zsun$ and its temperature
  1~keV, the implied radiative cooling time is 2~Gyr. If the galaxy is
  travelling around the sound speed relative to the cluster ($\sim
  1500 \kmps$), it would traverse the length in a tail in around
  20~Myr. Therefore radiative cooling is not an important physical
  process here.

The X-ray emission from the galaxy itself is consistent with a thermal
spectrum with a $0.77 \pm 0.09$ keV temperature. In these fits the
absorbing column density and metallicity were free, although the
metallicity is poorly constrained and the best fitting column density
is zero. The spectrum has 176 counts above the background and the
bolometric luminosity of the model is $4.5 \times 10^{39} \ergps$ and
the flux is $2 \times 10^{-15} \ergpcmsqps$.

\section{Discussion}
\subsection{Central cavity or cavities}
In clusters of galaxies, cavities are routinely observed in the
intracluster medium (for reviews see \citealt{McNamaraNulsen07} and
\citealt{Fabian12}). These cavities are inflated by jets, filled with
bubbles of relativistic plasma and are likely to combat cooling as
they do work on their surroundings. In the coronae of galaxies, radio
jets are found to be anti-correlated with the X-ray emission, as seen
in several objects besides NGC 4874, e.g. NGC 1265 \citep{Sun05_N1265}
and ESO 137-006 \citep{JonesMcAdam96}. In NGC 1265 there are small
X-ray indentations east and west of the nucleus where the jets leave
the coronal gas.

The best fitting model for the density distribution in NGC \,4889 has
a cavity or no X-ray emission inside 0.6~kpc radius. The residuals in
Fig.~\ref{fig:ngc4889residuals} do not obviously show the presence of
two cavities, as usually found in clusters. These features are usually
found in pairs, similarly to the radio jets they are created by. In
NGC 4889 the cavities, if present, are likely aligned along the line
of sight. An alternative possibility is that instead of X-ray
cavities, the central X-ray depression is due to a shock generated
from an outburst \citep[e.g.][]{Heinz98}. However, we do not observe
any central temperature increase associated with a shock.  If the
central cavity (or cavities; from here on we shall refer to a single
cavity for simplicity) is filled by relativistic material
($\gamma=4/3$), its enthalpy is $4PV$, where $P$ is the pressure (here
assumed to be the external pressure) and $V$ is its volume. Taking a
density of $0.084\pcmcu$ and temperature of $1.82\keV$, we estimate
that the cavity enthalpy is $5 \times 10^{55} \erg$.

The timescale for a bubble of radius $r$ takes to rise to a radius $R$
at its buoyancy velocity is $R / (0.5 v_K \sqrt{r/R})$
\citep{Churazov00}, where $v_K$ is the Keplerian velocity. Taking a
velocity dispersion of $\sim 380 \kmps$ \citep{McConnell11}, the
buoyancy timescale for a $0.3 \kpc$ radius bubble is 1 or 2~Myr, using
$R=0.3$ and 0.6 kpc, respectively. $R$ may be larger than this,
although it cannot be so large that it displaces the bubbles away from
the central peak in emissivity and reduces the flattening in the
surface brightness profile.

Taking the enthalpy and the 2~Myr timescale, the power needed to
inflate the cavity would be $8 \times 10^{41} \ergps$.  Therefore
during the period of activity the central black hole was supplying
energy at a rate one order of magnitude greater than that lost by
X-ray emission.  The scatter in relations of mechanical heating power
versus X-ray luminosity are of the order of one order of magnitude
\citep[e.g.][]{Birzan04, DunnFabian06} and so this outburst is not too
dissimilar to that found in clusters. The enthalpy of the bubble is
about an order of magnitude lower than the thermal energy of the
corona ($\sim 5 \times 10^{56} \erg$ within 3~kpc), so the total
energy accumulated during current 1-2 Myr long outburst is
insufficient to disrupt the corona.  The energy in the outburst may
also go into lifting the coronal gas in the gravitational
potential. $5 \times 10^{55} \erg$ would lift the corona by $\sim
0.1$~kpc, using the gravitational acceleration at 3~kpc radius.

As noted above, in most coronae the radio jets and lobes are
anti-correlated with the X-ray gas. This may be due to pressure
differentials between the external medium and the relativistic
material \citep{HeinzChurazov05}. Inside the jets/lobes the pressure
will be constant due to the high sound speed. However, in the corona
there will be a radial pressure gradient, where the pressure declines
with radius and the profile flattens outside in the ICM. There might
be an effect similar to what occurs when the end of a toothpaste tube
is squeezed; the radio channel will be closed by the coronal pressure
and the relativistic material displaced to larger radius into the ICM.

\subsection{Bondi accretion}
The black hole in NGC 4889 is one of the most massive known, with a
mass around $2 \times 10^{10} \Msun$ \citep{McConnell11}. Both NGC
4874 and 4889 show little evidence for dust in the HST F475W
data. Additionally, examining the Herschel 70 to 350 micron bands does
not show emission which would be associated with significant amounts
of cold dust. It is unlikely that any central X-ray surface brightness
depression in NGC\,4889 could be the result of obscuration. Therefore
hot accretion is a plausible fuelling mechanism for the massive black
hole in NGC\,4889. For such a massive black hole it may be possible to
resolve the material close at the Bondi radius. However, if there is a
cavity, then much of this material is likely to have been driven out.

The innermost temperature of the corona is around 1.9~keV. The Bondi
radius, $r_\mathrm{B}=2GM/c_\mathrm{s}^2$, for a black hole of mass
$M$ and a sound speed of $c_\mathrm{s}$, would be around 0.3~kpc,
close to the cavity radius. The Bondi accretion rate depends on the
density at the Bondi radius, which might be zero due to the presence
of the cavity. However, if we take the innermost value of the entropy
($\sim 9 \keV\psqcm$) and use the relation in \cite{Sun09}, this
implies the Bondi accretion rate is $0.2 \Msunpyr$. This translates
into a heating rate of $10^{45} \ergps$ at 10 per cent efficiency,
which is around four orders of magnitude greater than the
  X-ray luminosity of the galaxy. Inefficient accretion flows have
been inferred in other objects, such as the very massive black hole in
NGC\,1277 \citep{Fabian_1277_13}, Sgr A$^*$ at the centre of our
galaxy \citep{Rees82,Narayan95} and M87 \citep{DiMatteo03}.  It is
possible that the density has been enhanced around the central cavity
if material has been swept up by the outburst.

The black hole scaling relation with velocity dispersion suggests a
black hole mass for NGC\,4874 of $1.7 \times 10^{9} \Msun$
\citep{McConnellMa13} using $\sigma=290 \kmps$ \citep{Thomas07},
although scaling relations predict a much smaller black hole mass for
NGC\,4889 than measured. Similarly, the Bondi accretion rate inferred
for NGC\,4874 is $0.01 \Msunpyr$, if the inner entropy is $2.2
\keV\psqcm$. This would translate into an accretion luminosity of $6
\times 10^{43} \ergps$. This is over two orders of magnitude greater
than the energy lost by the X-ray emission of the galaxy. In NGC\,4874
the cooling rate in the absence of heating should be $\sim 0.3
\Msunpyr$.

NGC\,4889 hosts a point-like radio source with a 5 GHz spectral
luminosity of $\sim 10^{28.0} \: \rm erg\:s^{-1}\:Hz^{-1}$
\citep{Birkinshaw85}, coincident with the centre of the galaxy. An
extended source at 1.4 GHz was also detected by
\cite{MillerComaRadio09} with $\sim 10^{28.1} \rm
erg\:s^{-1}\:Hz^{-1}$.  If we use the most likely black hole mass
($2.1 \times 10^{10} \Msun$), the black hole activity fundamental
plane \citep{Merloni03} suggests that the X-ray luminosity of the
nucleus should be $\sim 2 \times 10^{37} \ergps$. This is well below
the upper limit ($\sim 10^{39} \ergps$) we obtain in Section
\ref{sect:fitting}. The inferred mechanical heating power from the
radio luminosity \citep{Birzan08} is around three times smaller than
calculated from the cavity enthalpy and timescale. This is well within
the large scatter in the mechanical power to radio luminosity
relation.  High resolution radio observations of NGC\,4889 should show
high frequency radio emission occupying the cavity at the centre.
NGC\,4874 has a core flux density at 5 GHz of $\sim 10^{28.1} \rm
erg\:s^{-1}\:Hz^{-1}$ \citep{Feretti87}.  Again, the X-ray luminosity
of the black hole as estimated from the fundamental plane ($6 \times
10^{38} \ergps$) is below the upper limit of $3\times 10^{39} \ergps$.

It appears that in NGC\,4889 that an AGN outburst is displacing the
X-ray emitting material in the centre. This is a clear case where the
AGN is affecting the corona directly. However, the case of NGC\,4874
appears to be the more common among coronal sources. The radio jets,
if present, are only visible outside the corona. In this case, the
jets appear to have passed their way through the corona before a
significant amount of their energy has been dissipated. However, the
core radio power of the two galaxies are very similar, which could
imply similar accretion rates \citep{Merloni03}.

\subsection{NGC 4889 coronal temperature}
The temperature of the NGC\,4889 corona is the highest known for a
galactic coronae \citep{Sun07,Sun09}. However the galaxy does have a
high velocity dispersion \citep{McConnell11}. \cite{Sun07} compare its
temperature with its velocity dispersion (using
$\beta_\mathrm{spec}$). Although there are other galaxies with similar
temperatures and velocity dispersions, it is the hottest of
those. There may be some heating due to the AGN activity in its
centre, although it depends on the stage of the outburst. The
gravitational potential of the very massive black hole can also affect
the gas pressure and temperature if it is in hydrostatic
equilibrium. We deprojected the X-ray surface brightness to fit the
observed spectral temperature profile assuming hydrostatic
equilibrium, fitting for the parameters of an NFW mass model and the
outer pressure \cite[the algorithm is detailed in][]{SandersRGS10}. If
the black hole potential is also included, we find that the
contribution to the gas temperature or pressure is only significant in
the inner $\sim 1.5$ arcsec radius. Therefore the mass of the black
hole is not a significant direct contributor to the high temperature
of the galactic corona.

\subsection{Stellar mass loss}
The X-ray gas masses for the galaxies within a radius of 3 kpc are
$1.2 \times 10^{8} \Msun$ and $9.4 \times 10^{7} \Msun$, for NGC 4874
and NGC 4889, respectively. If we adopt a value for the stellar mass
loss rate of $2.1 \times 10^{-12} \Msunpyr (L_K/L_{K,\odot})^{-1}$, we
estimate that the stellar mass loss rate for NGC\,4889 is
$1.7\Msunpyr$.  Therefore this could replace the material within the
X-ray corona in 54~Myr, although the X-ray corona (using a radius of 3
kpc) lies only within the inner $1/3$ of the optical light from the
galaxy. This could reduce the mass loss rate able to contribute to the
corona to $0.6\Msunpyr$ and the replacement timescale to
0.2~Gyr. However, if the stellar mass loss outside the corona were to
form overdense blobs, it might be possible for these to fall towards
the centre of the galaxy and contribute to the replacement of the
corona. Stellar mass loss may also not mix in a simple way with
the coronal material \citep{Panagoulia13}. The K band luminosity of
NGC\,4874 is around $2/3$ that of NGC\,4889, its coronal gas mass is
30 per cent greater and $1/4$ of the optical light is within 3 kpc,
so the mass loss rate is around $0.3 \Msunpyr$ and the replacement
timescale is 0.4~Gyr.

These resupply timescales can be compared to the mean radiative
cooling times in the centre of NGC\,4874 and NGC\,4889 of 0.04 and
0.2~Gyr, respectively, computed using X-ray surface brightness
deprojection, including the black hole mass for NGC\,4889. We
calculate a mass deposition rate, the cooling rate in the absence of
heating, of $\sim 0.3$ and $0.15 \Msunpyr$ for NGC 4874 and NGC 4889,
respectively, within a radius of a few arcsec, including the effect of
the gravitational potential. If the coronae are radiatively cooling,
stellar mass loss would be sufficient to replace the material lost by
cooling.

In NGC\,4889 the kinetic power of stellar mass loss,
$(3/2)\dot{M}_\star \sigma_\star^2 \sim 8\times 10^{40} \ergps$, is
approximately sufficient to replace the energy lost by X-ray
emission. In NGC\,4874, the lower stellar mass loss rate and velocity
dispersion ($290$ vs $380 \kmps$), give a kinetic power around five
times lower than the X-ray luminosity. \cite{Sun07} find that in their
sample of coronae the stellar mass loss kinetic power is on average a factor of
3.5 times lower than the X-ray luminosity, indicating further sources
of heat are required if mass loss from outside the coronal region does
not affect the corona.

Both galaxies show the presence of star formation. Using the
\emph{WISE} band 4, we obtain star formation rates of $\sim 0.1-0.2
\Msunpyr$ using the relation of \cite{Lee13} with a Kroupa initial
mass function. If we combine \emph{GALEX} FUV and \emph{WISE} fluxes,
using the relation of \cite{Hao11}, we obtain star formation rates of
$\sim 0.3 \Msunpyr$. The inferred star formation rates are close to
the $\sim 0.2 \Msunpyr$ cooling rates expected in the absence of
heating. If the black hole in NGC\,4874 is depositing its energy
outside of the corona, it may be ineffective in preventing the cooling
taking place. The relatively high star formation rate in
  NGC\,4889 is at odds with the current energy input of the AGN. This
  could be reconciled if the timescale for the evolution of the
  bubbles was underestimated or if the AGN input was intermittent.

\section*{Acknowledgements}
We thank A. Vikhlinin for helpful discussions about the gain
variation.  Support for this work was provided by the National
Aeronautics and Space Administration through \emph{Chandra} Award
Number GO2-13145X issued by the \emph{Chandra X-ray Observatory}
Center, which is operated by the Smithsonian Astrophysical Observatory
for and on behalf of the National Aeronautics Space Administration
under contract NAS8-03060. The scientific results reported in this
article are based on observations made by the \emph{Chandra X-ray
  Observatory} and data obtained from the \emph{Chandra} Data Archive.

\bibliographystyle{mn2e}
\bibliography{refs}

\end{document}